# On-chip optical isolator and nonreciprocal parity-time symmetry induced by stimulated Brillouin scattering

Jiyang Ma[1], Jianming Wen[2], Yong Hu[1], Shulin Ding[1], Xiaoshun Jiang[1*], Liang Jiang[3], and Min Xiao[1,4*]

[1]National Laboratory of Solid State Microstructures, College of Engineering and Applied Sciences, and School of Physics, Nanjing University, Nanjing 210093, China

[2]Department of Physics, Kennesaw State University, Marietta, Georgia 30060, USA

[3]Department of Applied Physics and Yale Quantum Institute, Yale University, New Haven, Connecticut 06511, USA.

[4]Department of Physics, University of Arkansas, Fayetteville, Arkansas 72701, USA

*emails: jxs@nju.edu.cn; mxiao@uark.edu.

Abstract    Realization of chip-scale nonreciprocal optics such as isolators and circulators is highly demanding for all-optical signal routing and protection with standard photonics foundry process[1]. Owing to the significant challenge for incorporating magneto-optical materials on chip, the exploration of magnetic-free alternatives has become exceedingly imperative in integrated photonics. Here, we demonstrate a chip-based, tunable all-optical isolator at the telecommunication band based upon bulk stimulated Brillouin scattering (SBS) in a high-Q silica microtoroid resonator. This device exhibits remarkable characteristics over most state-of-the-art implements, including high isolation ratio, no insertion loss, and large working power range. Thanks to the guided acoustic wave and accompanying momentum-conservation condition, SBS also enables us to realize the first nonreciprocal parity-time symmetry in two directly-coupled microresonators. The breach of time-reversal symmetry further makes the design a versatile arena for developing many formidable ultra-compact devices such as unidirectional single-mode Brillouin lasers and supersensitive photonic sensors.

Integrated photonic circuits demand dynamic optical isolation for advanced signal processing and communications. Because of the time-reversal symmetry retained in light-matter interaction, unfortunately, light wave transport in any linear, time-invariant optical system complies with the Lorentz reciprocity[2]. To violate such reciprocity and obtain asymmetric transmission, it essentially requires to break the time-reversal symmetry. In optics this is typically achievable by employing the magneto-optic Faraday effect. Despite its commercial success, such a well-established approach poses a severe challenge[3,4] for the incorporation with chip-scale photonics due to fabrication complexity with the mature complementary metal-oxide semiconductor (CMOS) technique and difficulty in locally confining magnetic fields as well as significant material losses. As a result, a vibrant search for different physical principles to obtain magnetic-free optical nonreciprocity has garnered a vast impetus especially in recent years. Alternative methods, most, as yet, far from practical realizations, resort to indirect interband transitions[5,6], optomechanical interactions[7-12], Kerr nonlinearities[13,14], gain/absorption saturation[15], thermo-optic effect[16], opto-acoustic interaction[17], Raman amplification[18], nonlinear parametric amplification[19,20], stimulated Brillouin scattering (SBS)[21], Bragg scattering[22], and mimicked nonlinear nonadiabatic quantum jumps[23]. Of these schemes, asymmetric optical transmission is mostly experimented with only injecting a light wave in either forward or backward direction but never both, except the works[11,14,19]. This type of implementation drawbacks was questioned in a recent theoretical proposal[24] by Shi et al, where they disproved Kerr or Kerr-type nonlinearities being capable of providing complete isolation due to dynamic reciprocity.

In parallel to the rapid progress on on-chip optical nonreciprocity, there has been an intense research interest in non-Hermitian photonic systems[25,26]. This originates from the observation made by Bender and Boettcher[27] that certain non-Hermitian Hamiltonians retaining the combined parity-time (PT) symmetry may have real spectra as long as a non-Hermitian parameter reaches above a specific threshold value. At threshold PT symmetry breaking spontaneously occurs, where the system transitions to a new phase associated with a pair of complex eigenvalues. Regardless of much theoretical success in the development of PT-symmetric quantum theory[28,29], it becomes highly challenging in search of such an elusive Hamiltonian in a real physical world. On the other hand, experiments with photonics can be intrinsically non-Hermitian due to the co-existence of gain and loss. Indeed, subsequent works[15,30-40] have undermined their relevance of quantum origin, and readily shown striking PT phase transitions in various optical settings by interleaving balanced gain and loss regions. The associated exceptional spectral properties thereupon translate into unique propagation and scattering behaviors for light, including power oscillations[31], unidirectional invisibility[32], coherent perfect laser absorbers[33], single-mode lasers[34,35], and supersensitive sensors[36,37].

The progresses with these two separate topics of chip-scale magnetic-free optical isolators and PT-symmetric optics put forward an interesting question: whether it is feasible to build a true optical isolator with PT symmetry[41]. This has initiated ongoing discussions with some gripping arguments that the realization of a PT-based optical isolator is indispensable of nonlinearities, in order to break the time-reversal symmetry. A close examination on these proposals reveals that the involved nonlinearities usually take the form of Kerr or Kerr-type, rendering them subject to the inevitable dynamic reciprocity. As such, complete isolation of backscattered signals is out of reach with those schemes. It thereby becomes fundamentally intriguing to know whether nonreciprocal PT symmetry could be attainable in other means.

Here we experimentally demonstrate, for the first time, a chip-based nonmagnetic optical isolator in a high-Q silica microtoroid resonator enabled by bulk SBS, in which coherent optically-driven acoustic waves can give rise to highly unidirectional photon-phonon interactions. Although SBS[42,43] is deemed to be the strongest of all optical nonlinear processes, exciting SBS in a chip-scale device is technically challenging because of the stringent requirements on materials and device geometry[44]. In contrast to previous cavity-optomechanical schemes based upon either surface SBS[10-12] or radial breathing mechanical mode[7-9], our platform remarkably exhibits many compelling features including larger isolation ratio, wider bandwidth, and much higher mechanical frequency. Essential for practical applications, the high frequency feature of the bulk SBS allows the pump light to be easily filtered from the signal field. By harvesting the directional momentum conservation inherent in the SBS process, we further report the first observation of nonreciprocal PT symmetry in two coupled microtoroid cavities with balanced gain and loss, where the SBS is exploited to produce direction-sensitive gain in the Brillouin cavity. Fully compatible with the current CMOS procedure, besides isolators and circulators, the structures could be configured for a number of demanding chip-based devices such as unidirectional single-mode Brillouin PT lasers and bi-scale ultrasensitive sensors.

The generation of SBS in a cavity requires a phase-matched three-mode system composed of two optical modes and one long-lived propagating acoustic mode in both frequency and momentum space[44], as schematically depicted in Fig. 1c, where their energies and momenta must simultaneously satisfy $\omega_1 - \omega_2 = \Omega_B$ and $k_1 - k_2 = \kappa_B$. Here $(\omega_{1(2)}, k_{1(2)})$ and $(\Omega_B, \kappa_B)$, respectively, denote the frequencies and wavenumbers of the high (low) energy optical mode and the propagating phonon mode. The coupling between these three modes is provided by the electrostrictive Brillouin scattering. Despite both forward- and backward-SBS phase matching can occur naturally in a microresonator, in this work a backward-SBS opto-acoustic interaction is particularly designed to take place within a high-Q microtoroid cavity via properly engineering its geometrical size (Fig. 1A). In experiment (Fig. 2A), a strong pump laser at frequency $\omega_p$ is launched to pump the high-energy optical mode $(\omega_1, k_1)$, while a weak tunable signal laser at frequency $\omega_s$ is, resonant with a resonator mode, employed to excite the low-energy optical mode $(\omega_2, k_2)$ instead. Consequently, the counter-propagating signal mode is expected to undergo strong resonant amplification due to the presence of Brillouin phase matching with the strong pump field and a longevous acoustic wave in the same medium. On the contrary, the co-propagating signal laser remains intact owing to the lack of the available Brillouin scattering. It is this unidirectional Brillouin gain that consents to the realization of a functioning optical isolator as well as nonreciprocal PT symmetry.

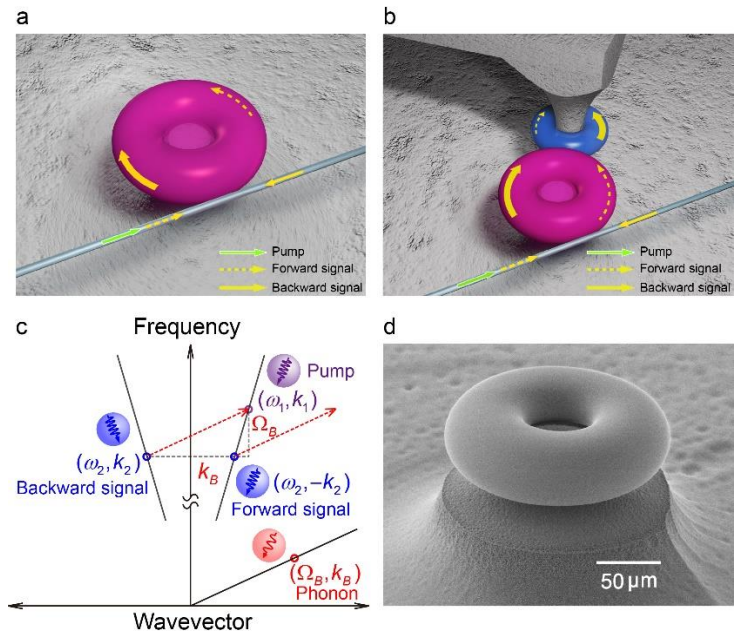

**Figure 1 On-chip optical isolator and nonreciprocal PT symmetry empowered by stimulated Brillouin scattering (SBS) with whispering-gallery-mode silica microtoroid resonators. a,** Schematic illustration of the experiment on optical isolation with a single toroid coupled to a tapered fiber. **b,** Schematic of the experiment on nonreciprocal $\mathcal{PT}$-symmetry with two toroids coupled with each other and the active one coupled to a tapered fiber. **c,** Schematic illustration of the requirements on phase and energy matching conditions for the SBS process. **d,** SEM image of the Brillouin toroid used in our experiments.

As schematically illustrated in Fig. 1a, our series of isolator experiments are performed simply with a Brillouin microresonator at room temperature and atmospheric pressure by simultaneously launching the signal lights from both forward and backward directions through a tapered fiber coupler for optical interface at 1550 nm. Figure 1d is an SEM image of the fabricated Brillouin microtoroid which has the principle and minor diameters of 175 μm and 54 μm, respectively. To identify the occurrence of SBS in the microtorid cavity, we first select two high-order optical transverse modes with separation matching the SBS frequency and optically pump the microcavity at the shorter-wavelength mode to above the lasing threshold. As shown in Fig. 2b, the Brillouin lasing indeed occurs at the longer-wavelength side. The loaded Q factors for the signal and pump waves are measured to be $1.40 \times 10^8$ and $4.36 \times 10^7$ with respect to their intrinsic Q factors of $1.99 \times 10^8$ and $1.04 \times 10^8$, respectively.

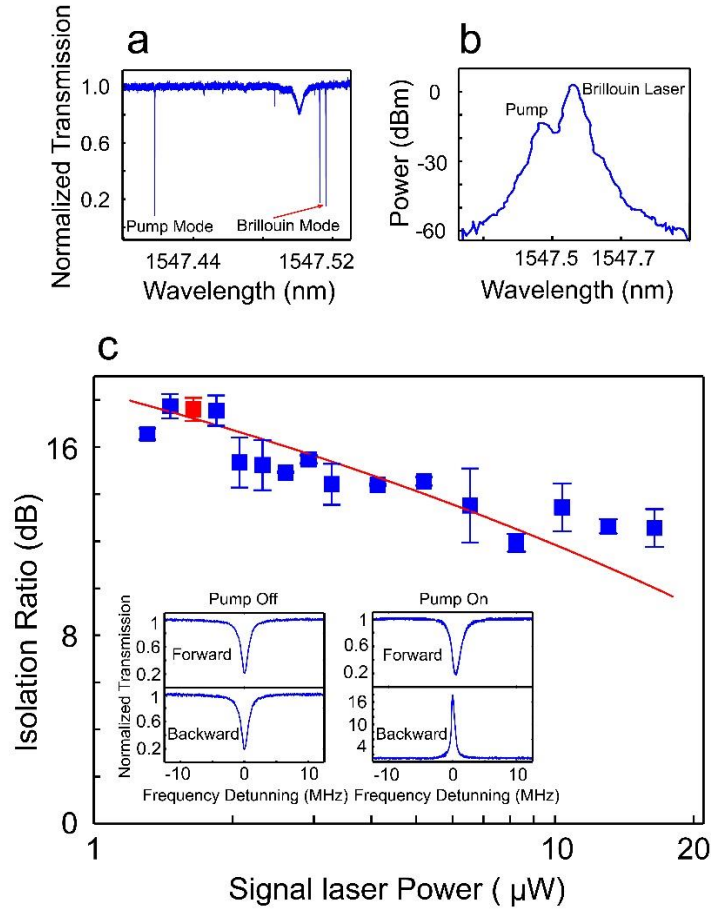

**Figure 2 Optical isolation versus input signal power. a,** Transmission spectrum when scanning the laser wavelength across the microtoroid shown in Fig. 1d. **b,** Brillouin lasing spectrum when the Brillouin microtoroid is pumped above the lasing threshold. **c,** Isolation ratio versus input signal laser power where the equal forward and backward signal laser powers were launched simultaneously from both directions. Inset: Typical transmission spectra corresponding to the red scatter in **c** The red line corresponds to the theoretical curve calculated through the method in supplementary information.

To test the isolation performance of our device, we start with the case of the same input signal powers from both directions but synchronously changing them. In such a case, the pump light is thermally locked to the shorter-wavelength cavity mode with a blue frequency detuning. During the measurement, the dropped pump power is fixed at 359.82 μW (below lasing threshold), and the coupling strength $\kappa$ between the cavity and tapered fiber is also fixed. The attained isolation ratio is depicted in Fig. 2C as a function of the input signal power. For the launched power range, the isolation ratio is well situated above 11.90 dB and even up to 17.73 dB for the bandwidth of 0.61 MHz. The reduction at higher input powers stems from the Brillouin gain saturation (Supplementary Note 3), similar to the reported gain saturation with doped erbium ions[15]. As a comparison, the insets display typical transmission spectra of the output signals as the pump turned on and off for forward and backward configurations. Without the pump, a Lorentzian dip profile is expected to be symmetrically observed in both directions in response to the cavity resonance. In contrast, when turning on the pump, the signal transport is dramatically altered: the forward input still remains the same absorption profile whereas the backward input experiences significant amplification. Notice that there is no insertion loss for this device since the signal light is amplified. In comparison with previous results, the SBS-induced optical isolation does reveal a number of advantages ranging from fast switching, great system stability, huge system parameter space, to low input signal power.

Thanks to the directional Brillouin gain, the Brillouin microcavity further empowers us to accomplish nonreciprocal PT symmetry by coupling with another passive microtoroid resonator with smaller size. The experimental setup is schematically depicted in Fig. 1b, where the inverted coupling arrangement[45] is adopted to ease the positioning adjustment. In the experiment, the two microtoroids are mounted on two nanopositioning stages for precise position-separation control. Moreover, we utilize two thermoelectric coolers (TECs) with a temperature stability of 2 mK to individually tune the cavity resonance wavelength via the thermo-optic effect so that both toroids are assured to always share the same resonant frequency during the measurements. The loaded Q factors of the Brillouin cavity are measured to be $3.16 \times 10^7$ at 1547.5 nm and $2.96 \times 10^7$ at 1547.4 nm, respectively, in contrast to their intrinsic Q-factors at $5.41 \times 10^7$ and $9.47 \times 10^7$. For the lossy microtoroid, its intrinsic Q factor is $9.72 \times 10^7$ at 1547.5 nm.

Since the Brillouin gain is only present in the backward configuration, the PT symmetry is investigated under balanced gain and loss for the backward signal input. While for the forward signal, the system is simply two passively coupled microcavities. To avoid the saturation nonlinearity, the dropped pump power is stabilized at 362.32 µW and the input (backward and forward) signal powers are maintained at a low power of 1.10 µW. In addition, the coupling strength $\kappa$ between the fiber and Brillouin cavity is fixed at $2\pi \times 2.55$ MHz. After carefully balancing the gain and loss in the backward direction, the evolutions of the transmitted signal spectra are presented in Fig. 3a by gradually decreasing the coupling strengths $\mu_{f,b}$ between two toroids while maintaining zero cavity frequency detunings. It is apparent that for the backward signal wave, frequency bifurcation induced by PT symmetry exhibits distinct features in spectral location change (Fig. 3c) and linewidth narrowing (Fig. 3b). In particular, a PT phase transition occurs at the exceptional point where $\mu_b = g = \gamma$. Theoretically, for $\mu_b > \gamma$ the system is in the unbroken phase and two real PT spectral branches $\omega_\pm$ with zero linewidth are quadratically displaced at $\pm\sqrt{\mu_b^2 - \gamma^2}$ away from the central cavity resonance frequency $\omega_0$. When $\mu_b < \gamma$, the PT symmetry spontaneously breaks down and the two spectral eigenvalues $\omega_\pm$ now become a complex conjugate pair. These behaviors have been well confirmed by the experimental results shown in Figs. 3b and 3c. Due to the spectral singularity of the complex optical potential, the significant signal amplification in the backward output (Fig. 3a) also verifies the theoretical prediction by Mostafazadeh[29]. For the forward signal mode, on the contrary, we anticipate to have least outputs at the two supermodes $\varpi_\pm = -i\frac{\gamma+\gamma_g}{2} \pm \sqrt{\mu_f^2 - \left(\frac{\gamma-\gamma_g}{2}\right)^2}$ (see Supplementary Note 1), which are determined by $\mu_f$ and the decay rates $(\gamma, \gamma_g)$ of the two microresonators. The measured output spectra in Fig. 3 show excellent agreements with the above theoretical analysis. As such, nonreciprocal PT symmetry is successfully demonstrated in this compound microcavity system. Given the tight connection of our device design to recently demonstrated single-mode lasers[34,35] and ultra-sensitive sensing[36,37], we expect to extend the applications to unidirectional single-mode Brillouin lasers and bi-scale supersensitive photonic sensors (see Supplementary Note 2) where nanoparticles with different sizes range could be distinguished from the forward and backward transmission spectra.

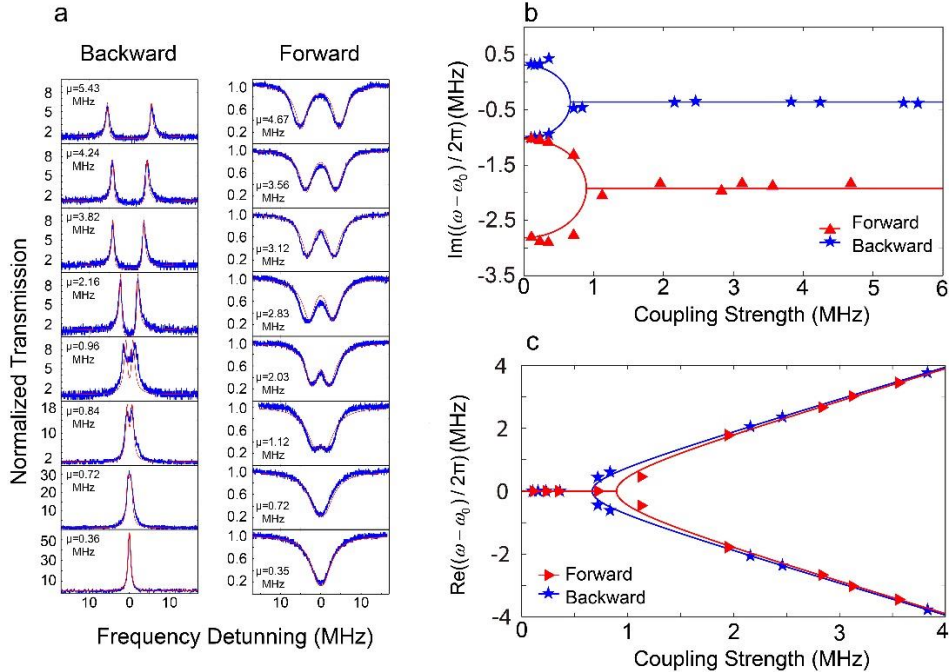

**Figure 3 Transmission spectra in nonreciprocal PT symmetry. a.** Output transmission spectra for forward and backward signal propagation configurations. μ represents the coupling strength between two cavities. **b & c.** The imaginary parts and real parts of the two eigenfrequencies of the two supermodes shown in **a** plotted as a function of the coupling strength for forward (red triangle) and backward (blue star) signal propagation configurations. Parameters: the dropped pump power is 362.32 $\mu$W and the coupling strength between the active cavity and the fiber is $2\pi \times 2.55$ MHz. (blue: experimental data; red: theoretical curves).

Unlike previous demonstrations, the SBS-induced optical nonreciprocity offers many appealing features beyond other alternatives. Despite the demonstrated isolation bandwidth is less than 1 MHz, it is believed to be extendable by orders of magnitude with further design optimization. From the device integration standpoint, the Brillouin microtoroid resonator has the advantages of compact footprint with CMOS compatibility and reliable isolation functionality, which are crucial in making nonreciprocal elements for on-chip photonic integration. As a fundamental building block, the structure allows a large operating range for the signal power from nanowatts to tens of microwatts. Furthermore, the introduction of nonreciprocal PT symmetry not only affirms the speculation on PT-assisted optical isolation, but also opens up new possibilities for constructing novel PT optical devices outperforming conventional facilities.

**References**

1. Jalas, D. *et al.* What is – and what is not – an optical isolator. *Nat. Photon.* **7**, 579-582 (2013).
2. Potton, R. J. Reciprocity in optics. *Rep. Prog. Phys.* **67**, 717-754 (2004).
3. Bi, L. *et al.* On-chip optical isolation in monolithically integrated non-reciprocal optical resonators. *Nat. Photon.* **5**, 758-762 (2011).
4. Stadler, B. J. H. & Mizumoto, T. Integrated magneto-optical materials and isolators: A review. *IEEE Photon. J.* **6**, 0600215 (2014).
5. Yu, Z. & Fan, S. Complete optical isolation created by indirect interband photonic transitions. *Nat. Photon.* **3**, 91-94 (2009).
6. Tzuang, L. D., Fang, K., Nussenzveig, P., Fan, S. & Lipson, M. Non-reciprocal phase shift induced by an effective magnetic flux for light. *Nat. Photon.* **8**, 701-705 (2014).
7. Shen, Z. *et al.* Experimental realization of optomechanically induced non-reciprocity. *Nat. Photon.* **10**, 657-661 (2016).
8. Ruesink, F., Miri, M.-A., Alu, A. & Verhagen, E. Nonreciprocity and magnetic-free isolation based on optomechanical interactions. *Nat. Commun.* **7**, 13662 (2016).
9. Fang, K. *et al.* Generalized non-reciprocity in an optomechanical circuit via synthetic magnetism and reservoir engineering. *Nat. Phys.* **13**, 465-471 (2017).
10. Kim, J., Kuzyk, M. C., Han, K., Wang, H. & Bahl, G. Non-reciprocal Brillouin scattering induced transparency. *Nat. Phys*. **11**, 275 (2015).
11. Kim, J., Kim, S. & Bahl, G. Complete linear optical isolation at the microscale with ultralow loss. *Sci. Rep.* **7**, 1647 (2017).
12. Dong, C. *et al*. Brillouin-scattering-induced transparency and non-reciprocal light storage. *Nat. Commun.* **6**, 6193 (2015).
13. Grigoriev, V. & Biancalana, F. Nonreciprocal switching thresholds in coupled nonlinear microcavities. *Opt. Lett.* **36**, 2131-2133 (2011).
14. Bino, L. D. *et al.* Microresonator isolator and circulators based on the intrinsic nonreciprocity of the Kerr effect. *Optica* **5**, 279 (2018).
15. Chang, L. *et al.* Parity-time symmetry and variable optical isolation in active-passive-coupled microresonators. *Nat. Photon.* **8**, 524-529 (2014).
16. Fan, L. *et al.* An all-silicon passive optical diode. *Science* **335**, 447-450 (2012).
17. Kang, M. S., Butsch, A. & Russel, P. St. J. Reconfigurable light-driven opto-acoustic isolators in photonic crystal fibre. *Nat. Photon.* **5**, 549-553 (2011).
18. Krause, M., Renner, H. & Brinkmeyer, E. Optical isolation in silicon waveguides based on nonreciprocal Raman amplification. *Electron. Lett.* **44**, 691-693 (2008).
19. Hua, S. *et al.* Demonstration of a chip-based optical isolator with parametric amplification. *Nat. Commun.* **7**, 13657 (2016).
20. Wang, K. *et al.* Non-reciprocal geometric phase in nonlinear frequency conversion. *Opt. Lett.* **42**, 1990-1993 (2017).
21. Huang, X. & Fan, S. Complete all-optical silica fiber isolator via stimulated Brillouin scattering. *J. Lightwave Technol.* **29**, 2267-2275 (2011).

22. Saha, K. *et al.* Chip-scale broadband optical isolation via Bragg scattering four-wave mixing *Conference on Lasers and Electro-Optics, paper QF1D.2* (Optical Society of America, 2013).
23. Choi, Y., Hahn, C., Yoon, J. W., Song, S. H. & Berini, P. Extremely broadband, on-chip optical nonreciprocity enabled by mimicking nonlinear anti-adiabatic quantum jumps near exceptional points. *Nat. Commun.* **8**, 14154 (2016).
24. Shi, Y., Yu, Z. & Fan, S. Limitations of nonlinear optical isolators due to dynamic reciprocity. *Nat. Photon.* **9**, 388-392 (2015).
25. Feng, L., El-Ganainy, R. & Ge, L. Non-Hermitian photonics based on parity-time symmetry. *Nat. Photon.* **11**, 752-762 (2017).
26. El-Ganainy, R. *et al.* Non-Hermitian physics and PT symmetry. *Nat. Phys.* **14**, 11-19 (2018).
27. Bender, C. M. & Boettcher, S. Real spectra in non-Hermitian Hamiltonians having PT symmetry. *Phys. Rev. Lett.* **80**, 5243-5246 (1998).
28. Bender, C. M. Making sense of non-Hermitian Hamiltonians. *Rep. Prog. Phys.* **70**, 947-1018 (2007).
29. Mostafazadeh, A. Pseudo-Hermitian representation of quantum mechanics. *Int. J. Geom. Methods Mod. Phys.* **7**, 1191-1306 (2010).
30. Peng, B. *et al.* Parity-time-symmetric whispering-gallery microcavities. *Nat. Phys.* **10**, 394-398 (2014).
31. Ruter, C. E. *et al.* Observation of parity-time symmetry in optics. *Nat. Phys.* **6**, 192-195 (2010).
32. Feng, L. *et al.* Experimental demonstration of a unidirectional reflectionless parity-time metamaterial at optical frequencies. *Nature Mat.* **12**, 108-113 (2013).
33. Wong, Z. J. *et al.* Lasing and anti-lasing in a single cavity. *Nat. Photon.* **10**, 796-801 (2016).
34. Feng, L., Wong, Z. J., Ma, R. M., Wang, W. & Zhang, X. Single-mode laser by parity-time symmetry breaking. *Science* **346**, 972-975 (2014).
35. Hodaei, H., Miri, M. A., Heinrich, M., Christodoulides, D. N. & Khajavikhan, M. Parity-time-symmetric microring lasers. *Science* **346**, 975-978 (2014).
36. Chen. W., Ozdemir, S. K., Zhao, G., Wiersig, J. & Yang, L. Exceptional points enhance sensing in an optical microcavity. *Nature* **548**, 192-196 (2017).
37. Hodaei, H. *et al.* Enhanced sensitivity at higher-order exceptional points. *Nature* **548**, 187-191 (2017).
38. Regensburger, A. *et al.* Parity-time synthetic photonic lattices. *Nature* **488**, 167-171 (2012).
39. Zhang, Z. *et al.* Observation of parity-time symmetry in optically induced atomic lattices. *Phys. Rev. Lett.* **117**, 123601 (2016).
40. Peng, P. *et al.* Anti-parity-time symmetry with flying atoms. *Nat. Phys.* **12**, 1139-1145 (2016).
41. Nazari, F. *et al.* Optical isolation via PT-symmetric nonlinear Fano resonances. *Opt. Express* **22**, 9574-9584 (2014).
42. Chiao, R. Y., Townes, C. H. & Stoicheff, B. P. Stimulated Brillouin scattering and coherent generation of intense hypersonic waves. *Phys. Rev. Lett.* **12**, 592-595 (1964).
43. Garmire, E. Perspectives on stimulated Brillouin scattering. *New J. Phys.* **19**, 011003 (2017).
44. Lee, H. *et al.* Chemically etched ultrahigh-Q wedge-resonator on a silicon chip. *Nat. Photon.* **6**, 369-373 (2012).

45. Li, G. Jiang, X. Hua, S. Qin, Y. & Xiao, M. Optomechanically tuned electromagnetically induced transparency-like effect in coupled optical microcavities. *Appl. Phys. Lett.* **109**, 261106 (2016)